\documentstyle[12pt,epsf]{article}

\newcommand{\eps}{\epsilon}
\newcommand{\ds}{\displaystyle}
\newcommand{\ra}{\rightarrow}
\newcommand{\be}{\begin{equation}}
\newcommand{\ee}{\end{equation}}
\newcommand{\bea}{\begin{eqnarray}}
\newcommand{\eea}{\end{eqnarray}}
\newcommand{\ci}{\cite}
\newcommand{\bi}{\bibitem}
\newcommand{\nono}{\nonumber \\}

\newcommand{\dd}{\partial}

\def\dal{\,\lower0.3ex\vbox{\hrule\hbox{\vrule\kern2pt\vbox{\kern4pt\kern4pt}
\kern2pt\vrule}\hrule}\,}

\begin{document}

\title{\sl Stationary finger solutions in the Hele-Shaw cell}
\vspace{1 true cm}
\author{G. K\"albermann, and R. Wallach\\Soil and Water department, 
Faculty of Agriculture, Rehovot 76100, Israel}
\date{}
\maketitle

\begin{abstract}

{\noindent}We solve numerically the nonlinear differential equation 
for the Hele-Shaw, Saffman-Taylor problem derived in the preceding work.
Stationary solutions with no free phenomenological parameters
are found to fit the measured patterns.
The calculated finger half-widths as a function of the physical
parameters of the cell, compare satisfactorily with experiment.
\end{abstract}

\newpage
\section{\label{solu}Stationary finger solutions in the Hele-Shaw cell}

In the preceding work\ci{k1}, we developed nonlinear
differential equations for the fingering phenomenon in the Hele-Shaw
cell.\ci{st}

The time independent equations for the stationary finger are

\bea\label{diff}
0&=&\eta_{xxx}-3~\frac{\eta_{xx}
^2~\eta_x}{1+\eta_x^2}~+~(1+\eta_x^2)^{\frac{3}{2}}~\frac{W(\eta)}{4~B}\nono
W(\eta)&=&\frac{4~\eps~\eta^2+1-5~\eta^2}{1-\eta^2}\nono
\epsilon&=&\frac{V}{U}
\eea
where the suffix indicates differentiation with respect to {\sl x}, $\ds \eta$
is the {\it y} coordinate of the finger profile.
$\ds \frac{1}{B}=\frac{12\mu~U~w^2}{{\tilde T}~b^2}$,
with $\ds \mu$ the viscosity of the displaced fluid, $\ds T$, the effective
surface tension and $\ds b$ the thickness of the Hele-Shaw cell.\ci{k1}
Equation (\ref{diff}), expressed in 
terms of the arclength measured from the tail of
the finger becomes

\bea\label{diffs}
0&=&\eta_{sss}+\frac{\eta_{ss}
^2~\eta_s}{1-\eta_s^2}+(1-\eta_s^2)~\frac{W(\eta)}{4~B}\nono
\eea

In eqs.(\ref{diff},\ref{diffs}), $\ds \eps=\frac{V}{U}$. With {\it U}
the velocity of the tip of the nose of the finger in the laboratory
(cell stationary) frame, and, {\it V} a velocity parameter determined
self-consistently from the existence of a stationary solution.
Asymptotically far back at the tail
we must have $W(\eta)~=~0$, or, 
$\ds \epsilon=\frac{5~\lambda^2-1}{4~\lambda^2}$.
The parameter $\epsilon$, is then determined by the solutions to
the equations and is not a phenomenological parameter.

Equation (\ref{diff}) can be transformed to a 
second order one in terms of the angle tangent to the curve, 
with $\eta$ obtained by integration. For $ds$ starting at the tail, 
where $\ds\theta\approx\pi$, $\ds ds~cos(\theta)=~-dx$.

Equation (\ref{diffs}) reads

\bea\label{diff2}
0&=&\frac{{\dd}^2\theta}{\dd s^2}-cos(\theta)~\frac{W(\eta)}{4~B}\nono
\eta&=&\lambda-\int{ds~sin(\theta)}
\eea

With $W(\eta)$ defined in eq.(\ref{diff}).

In the preceding work we analysed qualitatively the
predicted finger half-widths $\lambda$, as a function
of the parameter B. We found $\eps=1$ 
in the limit of $B\ra\infty$.
In this limit the potential determines the finger half-width 
to be $\lambda=1$, as expected.
At the other end of $B\ra~0$, or infinite viscosity limit, we argued\ci{k1}, 
that $\eps=0$ in the spirit of the no-slip condition for viscous fluids.
In this limit the potential $\ds W(\eta)$, yields
$\lambda=\frac{1}{\sqrt{5}}\approx 0.447$. Figure 8 of Tabeling et al.\ci{tab},
shows that for $\ds \frac{1}{B}\ra\infty$ 
the finger half-width for various aspect ratios tends 
to $\lambda\approx~0.45$.

Therefore we can be quite confident on the potential {\sl W}.
On the other hand, the dynamics as described
by the third order differential equation linear in the parameter {\sl B}
is certainly limited to low capillary numbers for which only the
lowest order Young-Laplace expression suffices. At larger capillary
numbers higher order terms in the curvature are needed.\ci{brow}
In the following we limit our calculations to relatively
low capillary numbers.

The numerical solution was started from positive $y=\lambda$ and followed 
around the finger to $y=-\lambda$. We employed
a fourth order Runge-Kutta algorithm.
We opted for the mixed
integrodifferential approach of eq.(\ref{diff2}), 
that proved easier to handle numerically.
 
For fixed {\sl B} we varied the value of the asymptotic half-width
(thereby fixing $\epsilon$), with $\theta'(s=)=0$ and 
$\theta(s=0)=180^0$, until a solution is found whose nose 
tip shows up at exactly $90^0$. 
Moreover, only solutions where the angle decreases smoothly from the
tail forwards were accepted. We rejected meandering (even slightly so)
solutions that could appear as a many pronged finger.

\begin{figure}
\epsffile{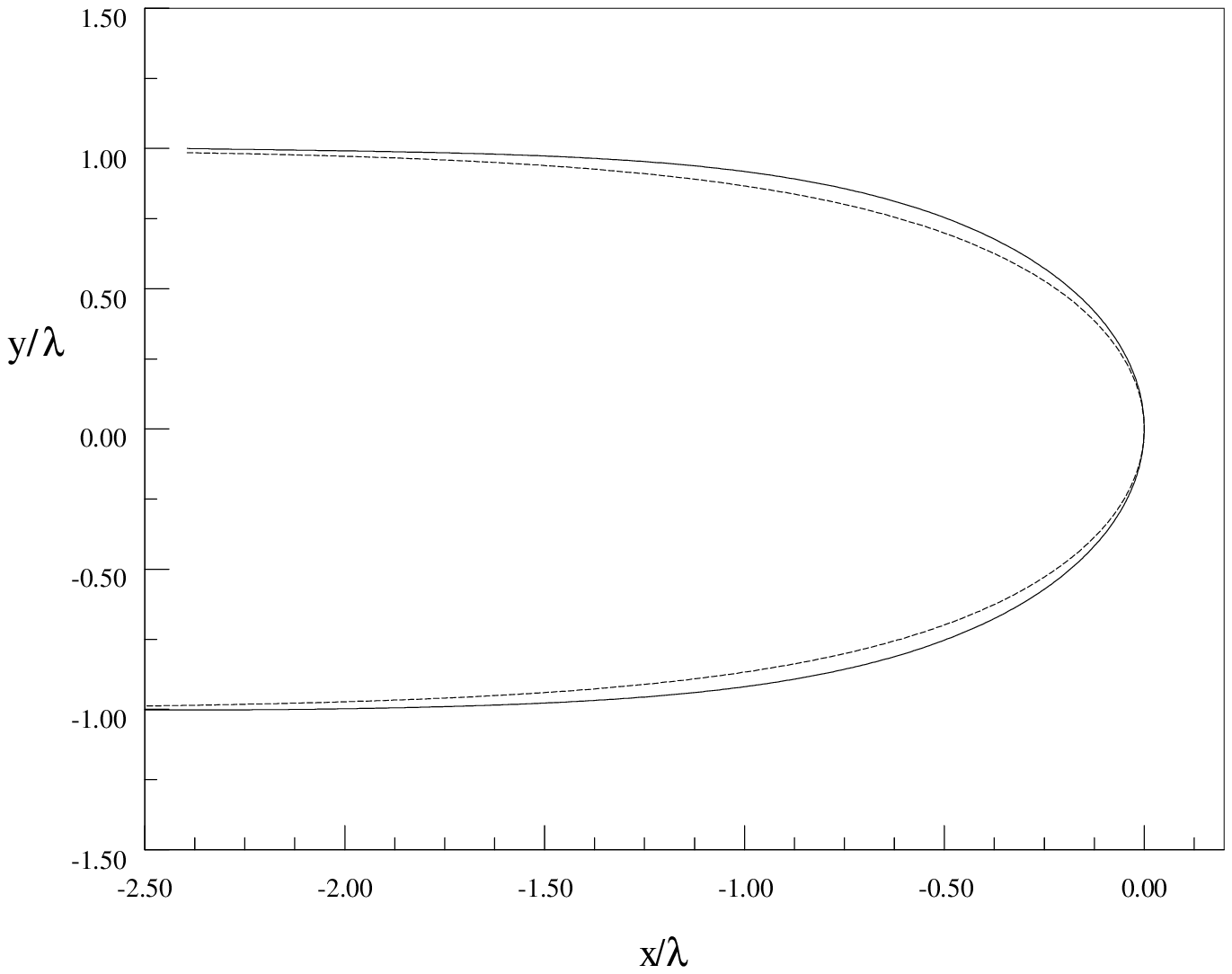}
\vsize=5 cm
\caption{\sl Rescaled finger profile for $\lambda=0.93$ at $\frac{1}{B}=12$. 
Numerical solution, solid line,
analytical solution of Pitts [10], dashed line.}
\label{fig2}
\end{figure}
We proceed to show 3 profiles found for finger half-widths 
$\lambda=0.93,0.82,0.66$. They are compared with the analytical solution 
of Pitts\ci{pit} that fits the data extremely well for finger half-widths
smaller than around $\lambda\approx 0.8$ and lies slightly 
inside the data points for higher
values of $\lambda$. The figures are depicted for
rescaled coordinates in terms of $\lambda$ as has became customary\ci{mc,pit}.
\begin{figure}
\epsffile{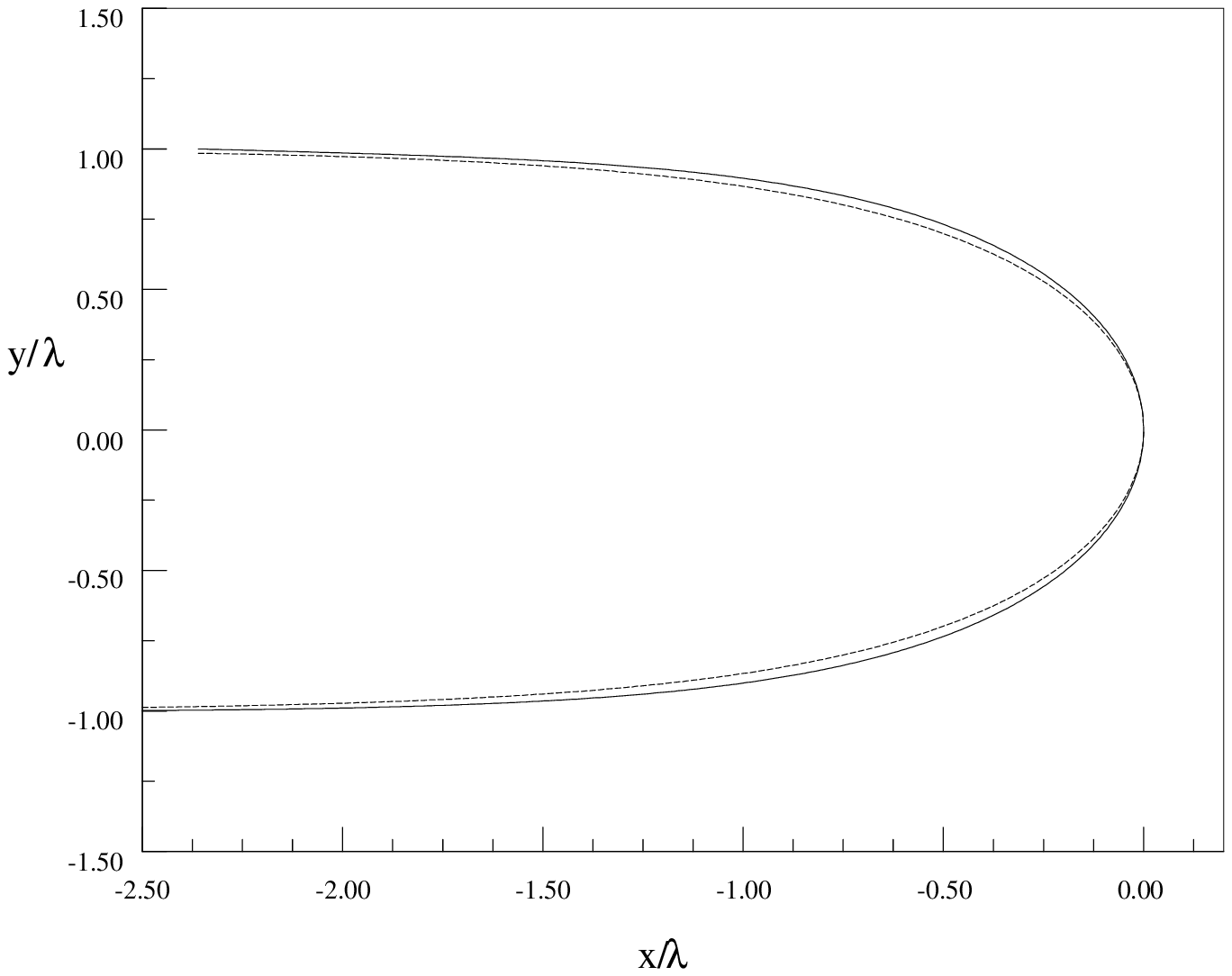}
\vsize=5 cm
\caption{\sl Rescaled finger profile for $\lambda=0.82$ at $\frac{1}{B}=20$. 
Numerical solution, solid line,
analytical solution of Pitts [10], dashed line.}
\label{fig3}
\end{figure}

From figures 2-4 it is clear that the solutions match the 
phenomenological solutions of Pitts $\ds cos(\frac{\pi}{2}\frac{~y}{\lambda})
e^{-\frac{\pi}{2}\frac{~x}{\lambda}}=1$, and are even superior to it for large
values of $\lambda$.
A graph of the asymptotic finger half-width $\lambda$ as a function of 
{\sl B}, is shown in figure 5.
The present results improve upon the calculations of
McLean and Saffman\ci{mc} for small capillary numbers.
Note however, that our results do not include finite gap corrections that
lift the curve upwards for small {\sl B}, 
whereas the results of McLean and Saffman\ci{mc}.
shown in the figure do adjust for such corrections. Without
these corrections the latter results fall much below the depicted curve.

\begin{figure}
\epsffile{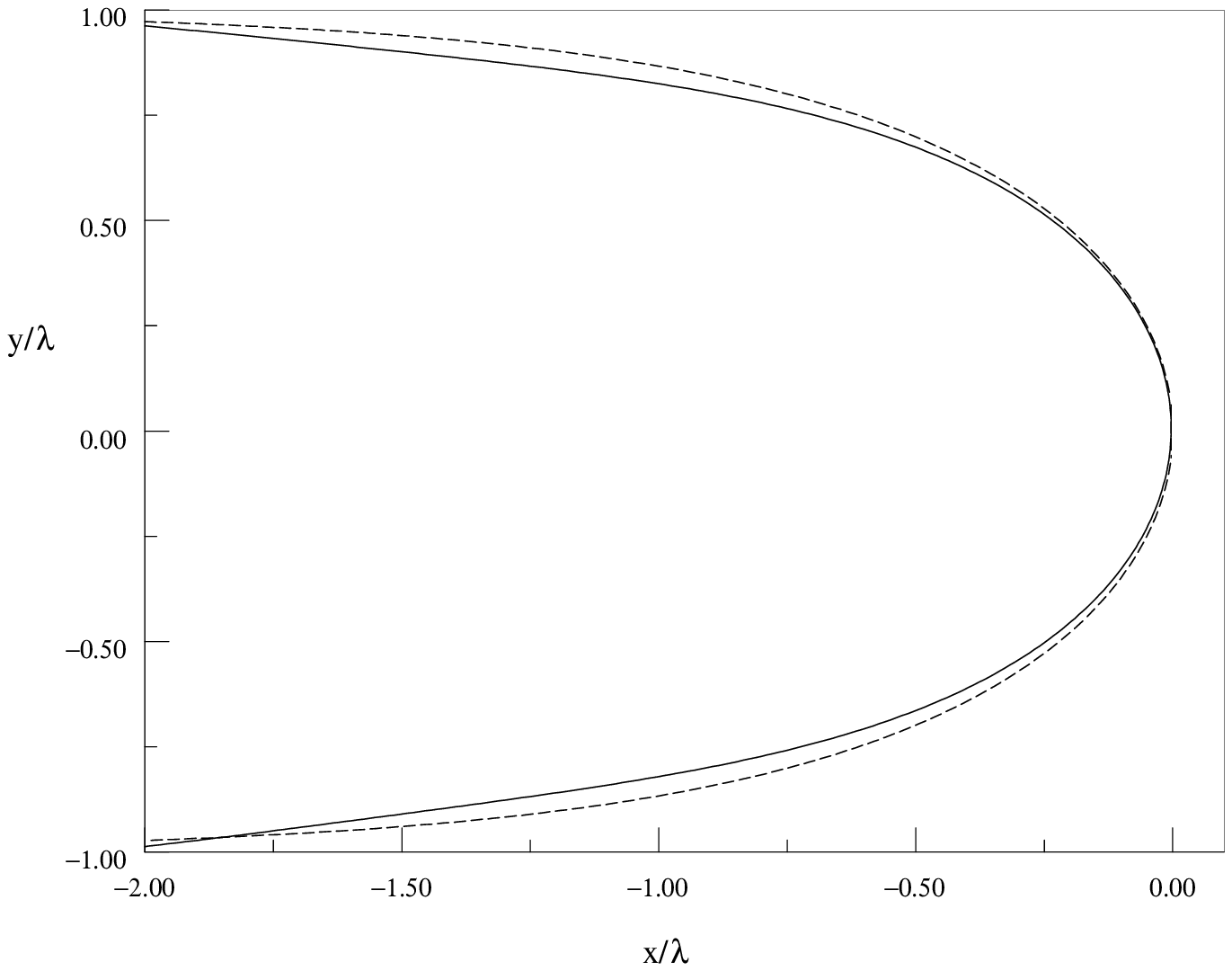}
\vsize=5 cm
\caption{\sl Rescaled finger profile for $\lambda=0.66$ at $\frac{1}{B}=38$. 
Numerical solution, solid line,
analytical solution of Pitts [10], dashed line.}
\label{fig4}
\end{figure}

\begin{figure}
\epsffile{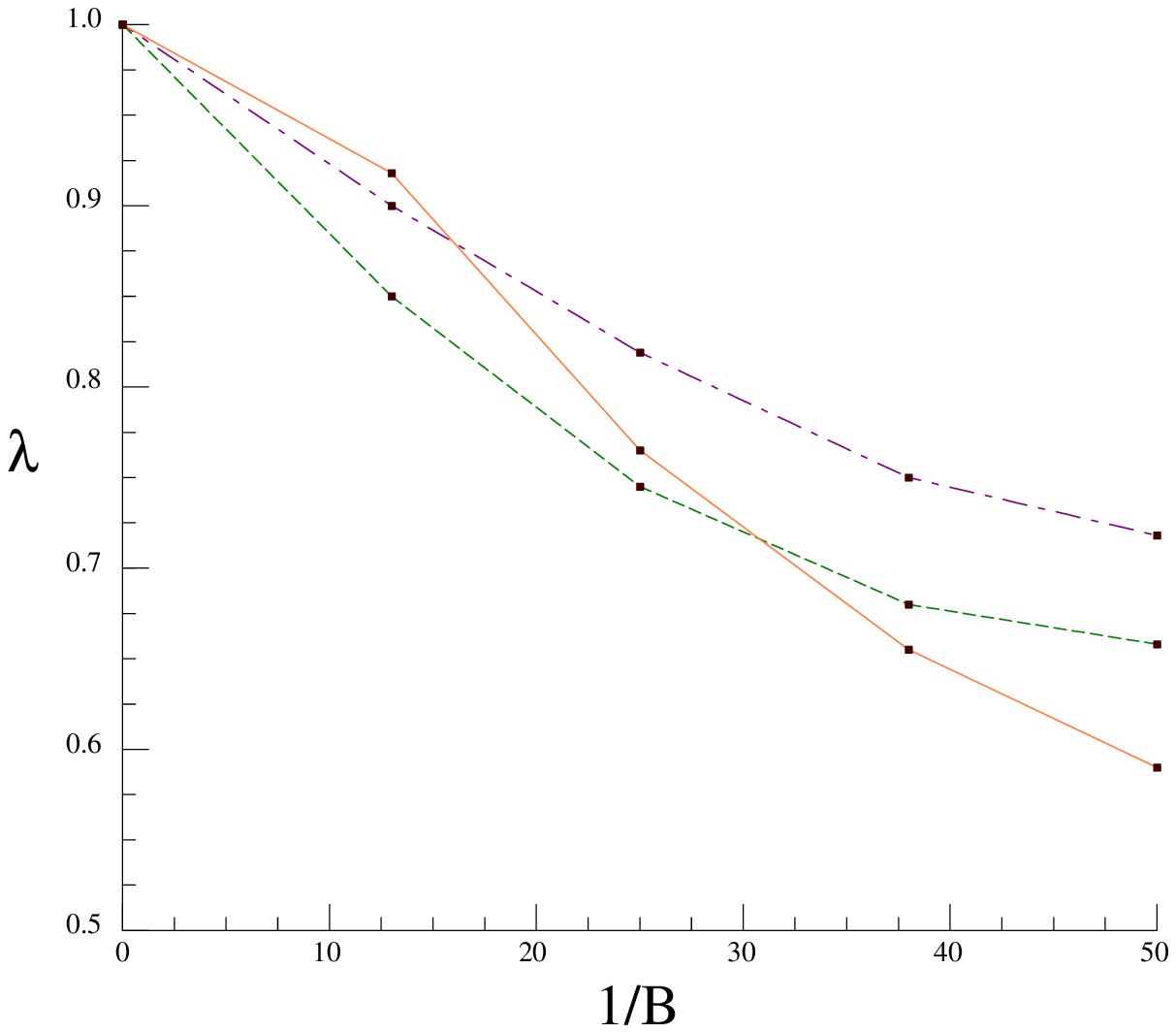}
\vsize=5 cm
\caption{\sl Asymptotic half-width of the finger, $\lambda$ as
a function of  $\frac{1}{B}$. Present work, full line,
theoretical results of McLean and Saffman [7] corrected for
finite film effects, dashed line, and,
 experimental results of Tabeling et al. [9], dash-dot line.}
\label{fig5}
\end{figure}

We found numerically stationary finger profiles for {\sl B} $<$ 50.
Beyond this value, the large contributions introduced by the potential in 
eq.(\ref{diff2}), render the integration from the tail unstable.
The curve of $\lambda$ versus $\ds \frac{1}{B}$ 
Several attempts were made in the present study, 
to improve the algorithm by matching solutions
run from the tail and from the nose. However, starting from the nose
is very problematic, because of the singularity in the slope.

We found that for a 
a fixed value of {\sl B}, there corresponds a single value of $\lambda$; a 
unique finger solution.
The selection of the finger half-width is univocous.
However, we cannot rule out completely the existence of
other unstable branches not accessed by the numerical method, 
as found by Vanden-Broek\ci{van} for the McLean and Saffman\ci{mc} set of
integrodifferential equations.

\section{Conclusion}

The numerical stationary profiles obtained from the
nonlinear differential equation derived in \ci{k1}, fit well
the experimental ones, that are, in turn, quite accurately
reproduced by the phenomenological solution of Pitts\ci{pit}.
The predicted finger half-width $\lambda$,
as a function of capillary number,
fits the experimental results in the low capillary number regime.
In the course of the numerical investigation we found solutions
that correspond to various multiple finger-like structures.
These structures originate from specific boundary conditions
at $y=\lambda$. 
The existence of such complicated patterns is quite
encouraging, especially if we want to proceed further to the production,
development, competition and splitting between fingers with
the time-dependent equation displayed above.

\end{document}